\documentclass[12pt, english]{article}
\usepackage[top = 2 cm, bottom = 3 cm, left = 2.5 cm, right = 1.8 cm]{geometry}
\usepackage{authblk, cite, color, amssymb, amsmath}
\usepackage[hypertex, colorlinks = true, linkcolor = blue, citecolor = red]{hyperref}
\usepackage[dvipsnames]{xcolor}
\usepackage{babel}

\begin{document}

\title{Hidden Conformal Symmetry of Smooth Braneworld Scenarios}

\author{\bf G. Alencar\footnote{e-mail: geova@fisica.ufc.br}}

\affil{\small Departamento de F\'{\i}sica, Universidade Federal do Cear\'{a}
Caixa Postal 6030, Campus do Pici, 60455-760, Fortaleza, Cear\'{a}, Brazil}

\author{\bf Merab Gogberashvili\footnote{e-mail: gogber@gmail.com}}

\affil{\small Javakhishvili Tbilisi State University, 3 Chavchavadze Avenue, Tbilisi 0179, Georgia \authorcr
Andronikashvili Institute of Physics, 6 Tamarashvili Street, Tbilisi 0177, Georgia}

\maketitle

\begin{abstract}
In this paper we generalize our previous model (arXiv: 1705.09331), on a hidden conformal symmetry of smooth braneworld scenarios, to the case with two real scalar fields non-minimally coupled to gravity. The gauge condition reduces the action of the system to the action were gravity minimally couples to one of the scalar fields, plus a cosmological constant. We show that, depending on the internal symmetry of the scalar fields, the two possibilities, $SO(2)$ or $SO(1,1)$, emerge. In the $SO(2)$ case we get a ghost-like scalar field action, which can describe two models -- Standing Wave and Sine-Gordon smooth braneworlds. For the $SO(1,1)$ case we get the standard sign for the kinetic part of the scalar field. By breaking the $SO(1,1)$ symmetry (but keeping the conformal one) we are able to get two Randall-Sundrum models, with a non-minimal coupling and with a scalar field having hyperbolic potential. We conclude that this method can be seen as a solution-generating technique and a natural way to introduce non-trivial scalar fields that can provide smooth braneworld models.


\end{abstract}
\vskip 5mm


\section{introduction}

Some models with large extra dimensions emerged about two decades ago \cite{ArkaniHamed:1998rs, Antoniadis:1998ig, Gogberashvili:1998vx, Gogberashvili:1998iu, Gogberashvili:1999tb, Gogberashvili:1999ad, Randall:1999ee, Randall:1999vf}. Unlike the Kaluza-Klein approach with compact extra dimensions, in these models some mechanism must confine the fields to the brane, a $(3+1)$-subspace of a higher dimensional theory. In \cite{ArkaniHamed:1998rs} a mechanism was considered that trap gauge fields, however it does not work for gravity. Soon a way to confine gravity was found by the introduction of a non-factorizable metric \cite{Gogberashvili:1998vx, Randall:1999vf}. In this model, called RS2, gravity is confined to the brane, however, other fields cannot be localized on the brane. In fact, only gravity and scalar fields are trapped and only gravity is not enough to trap spin-$1$ and spin-$1/2$ fields on the brane \cite{Bajc:1999mh} (see \cite{Liu:2017gcn} for a review). Later, by considering a non-minimal coupling with gravity, the mechanism of gauge fields localization in the RS2 scenario has been found, with the respective phenomenological prevision of a small cosmological photon mass \cite{Alencar:2014moa, Zhao:2014iqa, Alencar:2015rtc}, however, the origin of this coupling was unexplained.

The solution to the above problems has been found recently using a hidden conformal symmetry of RS2 model \cite{Alencar:2017dqb}, where non-minimal coupling emerges naturally. By considering conformal torsion it was also obtained the universal localization of fermion fields (Note that this kind of strategy has been used to explain chaotic inflation models \cite{Kallosh:2013wya, Kallosh:2013xya}). This hidden symmetry also was found in RS1 models and some consequences to cosmology was considered \cite{McFadden:2004se}.


\section{Setup of the Model}

The model \cite{Alencar:2017dqb} is based on the 5D action
\begin{equation} \label{L_G}
L_G = \xi\chi^2R - \frac 12 \chi\nabla^2 \chi - u \chi^n - \mu \chi^n\delta(z)~,
\end{equation}
where $R$ represents 5D Ricci scalar, $\chi$ is a real scalar field, $z$ denotes the extra spatial coordinate and $\xi$, $u$, $\mu$ and $n$ are some constants. This Lagrangian is invariant under the conformal transformations
\begin{equation}
\tilde{g}{}_{MN} = e^{2\rho} g_{MN}~, \qquad \tilde{\chi} = e^{-3\rho/2}\chi~,
\end{equation}
if
\begin{equation}
\xi = \frac {3}{32}~, \qquad n = \frac {10}{3}~.
\end{equation}

Conformal symmetry can be used to describe the system in two different ways:
\begin{itemize}
\item {By fixing the scalar field to a constant,
\begin{equation}
\chi = \chi_0~,
\end{equation}
we get
\begin{equation}
L_G = \xi\chi_0^2 R - u \chi_0^n - \mu\chi_0^n\delta(z) ~.
\end{equation}
We see that the breaking of conformal symmetry fixes the energy scale for gravity by choosing
\begin{equation}
\xi \chi_0^2 = 2 M^3~.
\end{equation}
With this we obtain the full RS2 model in the standard form. We are left with the free parameter $u$, which can provide the cosmological constant if we choose
\begin{equation}
u\chi_0^n = 2\Lambda~.
\end{equation}
One solution to this is given by the 5D metric,
\begin{equation}
ds^2 = e^{2A(z)}\eta_{MN}dx^Mdx^N ~,
\end{equation}
with \cite{Randall:1999vf}
\begin{equation}
A(z) = -\ln(k|z|+1)~; \qquad \mu = 24 M^3 k~; \qquad \Lambda = -24M^3 k^2~.
\end{equation}
}
\item{The second way of describing the system (\ref{L_G}) is by fixing the warp factor to
\begin{equation}
e^{A(z)} = 1 ~.
\end{equation}
Then we get the Lagrangian
\begin{equation}
L_G = - \frac 12 \chi \chi'' - u \chi^n - \mu \chi^n \delta(z) ~,
\end{equation}
which leads to the 5D equation of motion:
\begin{equation}
\chi'' + nu \chi^{n-1} + n\mu\chi^{n-1}\delta(z) = 0 ~.
\end{equation}
Here primes denote derivatives with the respect to the extra coordinate $z$. The solution to the above equation is
\begin{equation}
\chi = \frac {C}{(k|z|+1)^{2/(n-2)}}~,
\end{equation}
with
\begin{equation}
C^{n-2} = \frac {9 k^2}{8u} ~, \qquad \mu = \frac {4u}{5k}~.
\end{equation}
The gravitational part of the action becomes,
\begin{equation}
S = \xi C\int\frac{dz}{(k|z|+1)^3}\int d^4xR_4
\end{equation}
and in order to recover 4D gravity we must impose the additional constraint
\begin{equation}
\xi C = 2M^3~.
\end{equation}
}
\end{itemize}
With this, all constants of the model agree in both gauges. We also should point that the integration factors are identical in both cases, giving the same relations between the constants. As said above, by using this as a guiding principle the mechanism of localization of matter fields have been found, including the universal localization of fermion fields \cite{Alencar:2017dqb}.

An important point that can be posed now is a smooth version of the above model. We look for a conformal model that can provide a kink-like solution to our scalar field equation, which recovers AdS for large $z$. Smooth solutions are important to avoid naked singularities on the brane and can be generated by adding a scalar field potential \cite{Kehagias:2000au, Barbosa-Cendejas:2013cxa, BarbosaCendejas:2005kn, Chinaglia:2016aat, HerreraAguilar:2010kt, Dzhunushaliev:2008zz, Dzhunushaliev:2009va}(for the cases without a scalar field potential see \cite{HerreraAguilar:2009wc}). These solutions provide a rich structure of resonances, or metastable massive modes over the membrane \cite{Landim:2011ki, Landim:2011ts, Alencar:2012en, Landim:2010pq, Li:2017dkw}. Smooth solutions with hyperbolic and trigonometric functions was considered in \cite{Gremm:1999pj, Bazeia:2017nlo, Bazeia:2012qh}. A smooth version of the RS model with a phantom scalar field, called Standing Wave Braneworld, has been found in \cite{Gogberashvili:2009yp, Gogberashvili:2010yf, Gogberashvili:2016vwt}. Also using a phantom scalar field the authors in \cite{Koley:2004au} found a smooth version by using a trigonometric potential. By a numerical study, in \cite{Farakos:2005hz, Farakos:2006tt} it is shown that models with non-minimal coupling have smooth and stable solutions. However, as far as we know, these unusual potentials, which also appear in models of conformal mechanics \cite{Holanda:2014zfa}, are introduced \emph{ad hoc}.

Coming back to the problem of generating smooth solutions for the model \cite{Alencar:2017dqb} we face a problem: conformal symmetry uniquely determines the potential and we are not allowed to introduce a new potential, such as $\lambda \chi^4$, hyperbolic or trigonometric potentials. Another problem is that the kinetic part of our scalar field has the wrong sign and behaves as a ghost.

As we mentioned above, we cannot add a $\lambda \chi^{4}$ potential to the Lagrangian (\ref{L_G}) since this breaks conformal invariance. Therefore, the only way to get some freedom in our model is to add one more conformal scalar field with the Lagrangian
\begin{equation}
L_{\chi_{2}} = \epsilon \left( \xi\chi_{2}^{2}R -\frac{1}{2} \chi_{2}\nabla^{2}\chi_{2}\right) ~,
\end{equation}
where $\epsilon = \pm 1$ is the sign function. Then (\ref{L_G}) takes the form:
\begin{equation} \label{generalaction}
L_G = \xi\left(\chi_1^2 + \epsilon \chi_2^2\right)R - \frac 12 \left(\chi_1\nabla^2\chi_1 + \epsilon \chi_2 \nabla^2 \chi_2\right) + U \left(\chi_1,\chi_2\right)~,
\end{equation}
where $U \left(\chi_1, \chi_2\right)$ is some potential term. This Lagrangian has a $SO(1,1)$ or $SO(2)$ symmetry depending on the sign of $\epsilon$ and can be used as a conformally symmetric extension of the model \cite{Alencar:2017dqb}. In order to preserve conformal invariance, the potential in (\ref{generalaction}) must have the form
\begin{equation}
U \sim \chi_{1}^{m}\chi_{2}^{n} \qquad \left(m + n = \frac {10}{3}\right)
\end{equation}
This new freedom is at the center of the strategy for obtaining smooth solutions.


\section{Hidden Symmetry of Phantom Models}

In this section we explore the ghost case with $\epsilon = +1$ and the Lagrangian
\begin{equation}\label{SO(2)}
L_G = \xi\left(\chi_1^2 + \chi_2^2\right)R - \frac 12 \left(\chi_1\nabla^2\chi_1 + \chi_2 \nabla^2 \chi_2\right) + U \left(\chi_1,\chi_2\right)~.
\end{equation}


\subsection{Standing Wave Braneworld}

The Lagrangian (\ref{SO(2)}) has a conformal symmetry and if $U=0$ the additional internal $SO(2)$ symmetry for the scalar fields $\chi_1$ and $\chi_2$. The potential that preserves these symmetries is given by
\begin{equation}
U = u \left(\chi_1^2 + \chi_2^2\right)^{5/3}~.
\end{equation}
We can use our conformal symmetry to choose the gauge
\begin{equation}\label{gaugeSO(2)}
\xi \left(\chi_1^2 + \chi_2^2\right) = 2M^3 ~.
\end{equation}
Then the gravitational part of (\ref{SO(2)}) reduces to the standard Einstein-Hilbert Lagrangian. We also will find that the potential will behave as a cosmological constant $U=\Lambda$. The constraint (\ref{gaugeSO(2)}) provides that the scalar field equations must have a solution given by:
\begin{equation}
\begin{split}\label{constraintSO(2)}
\chi_1 &= \sqrt{\frac{2M^3}{\xi}} ~\cos \left(\sqrt{\frac{\xi}{2M^3}}\phi \right)~,\\
\chi_2 &= \sqrt{\frac{2M^3}{\xi}} ~\sin \left(\sqrt{\frac{\xi}{2M^3}}\phi \right)~.
\end{split}
\end{equation}
With this the Lagrangian becomes
\begin{equation}
L_G = 2M^3 R - \frac 12 \phi\nabla^2 \phi - \Lambda ~.
\end{equation}
This Lagrangian (\ref{SO(2)}) is exactly the one used for the Standing Wave Braneworld \cite{Gogberashvili:2009yp, Gogberashvili:2010yf, Gogberashvili:2016vwt}. Therefore, it can be said that this model has a conformal origin.


\subsection{Negative Tension Sine-Gordon Model}

Now we consider a conformal model which breaks the $SO(2)$ symmetry. The conformal symmetry will be preserved if we change our potential $U$ by a dimensionless function $V(\chi_1, \chi_2)$, namely
\begin{equation}\label{breakSO(2)}
\tilde{U} = u \left(\chi_1^2 + \chi_2^2 \right)^{5/3} V \left(\chi_1,\chi_2 \right)~.
\end{equation}
If $V=1$ we will recover our symmetry. Now when we fix the gauge (\ref{gaugeSO(2)}) and use the solution (\ref{constraintSO(2)}), our potential (\ref{breakSO(2)}) reduces to the effective potential
\begin{equation}
\tilde{U} = \Lambda V \left(\sin \sqrt{\frac{\xi}{2M^3}}\phi, \cos \sqrt{\frac{\xi}{2M^3}}\phi \right)~,
\end{equation}
where $\phi$ is a real scalar field and $V$ is a dimensionless function of $\phi$. An important fact about this model is that this kind of potential is generated naturally by fixing the energy scale of gravity in 5D.

Since the function $V(\sin \phi, \cos \phi)$ is arbitrary, we can generate from it many trigonometric potentials commonly used in the literature, for example the Sine-Gordon model. A smooth brane generated by a Sine-Gordon potential with a ghost-like scalar field has been found in \cite{Koley:2004au}, we will reconstruct it here by using the superpotential method.

The equations of motions for the scalar-gravity system above can be reduced to
\begin{equation}
\begin{split}\label{coupledeqSO(2)}
- \frac 12 \phi'^2 - V(\phi) - \Lambda &= 24M^3 A'^2 ~, \\
- \frac 12 \phi'^2 + V(\phi) + \Lambda &= - 12M^3A'' - 24M^3 A'^2 ~.
\end{split}
\end{equation}
In these equations and from now on, as in \cite{Kehagias:2000au} we are using
\begin{equation}
ds^2 = e^{2A} \eta_{\mu\nu}dx^{\mu}dx^{\nu} + dz^2 ~.
\end{equation}

A solution to the system (\ref{coupledeqSO(2)}) can be found by introducing a superpotential $W$, such that the potential can be written as
\begin{equation}
U(\phi) = - \frac 12 \left(\frac{\partial W}{\partial\phi}\right)^2 - \frac{1}{6M^3}W^2 ~.
\end{equation}
We see that
\begin{equation}
\phi' = \frac {\partial W}{\partial\phi}~, \qquad A' = \frac {W}{12M^3}
\end{equation}
are solutions of the system (\ref{coupledeqSO(2)}). Since we want to obtain the Sine-Gordon model we take the superpotential
\begin{equation}
W = - \sqrt{6\Lambda M^3} ~\sin \left(\sqrt{\frac{\xi}{8M^3}}\phi\right) ~,
\end{equation}
where $\Lambda$ is a dimensionless parameter. This in fact provides
\begin{equation}
U(\phi) = 3\Lambda \left[- \left( \frac 16 + \frac{\xi}{16}\right) + \left(\frac 16 - \frac{\xi}{16}\right) \cos \left( \sqrt{\frac{\xi}{2M^3}}\phi \right)\right]~,
\end{equation}
which has the desired form. Then the solution to the system (\ref{coupledeqSO(2)}) is
\begin{equation}
\begin{split}\label{solution}
\phi =& \sqrt{\frac{32M^3}{\xi}} ~\arctan \left(\tanh \sqrt{\frac{6\Lambda \xi^2}{M^3}} z\right) ~,\\
A =& \frac {1}{12\xi} ~\ln \left(2\cosh \sqrt{\frac{6\Lambda \xi^2}{M^3}} z \right) ~.
\end{split}
\end{equation}
When $z \to \infty$ we have $A = kz$, with $\Lambda = 24M^3 k^2$, as desired. Similar solution has been used in \cite{Koley:2004au} in order to localize fermion fields. Consequently, our objective is reached.


\section{Hidden Symmetry of Real Scalar Models}

Now we consider the case $\epsilon = -1$ in (\ref{generalaction}) to get
\begin{equation}\label{SO(1,1)}
L_G = \xi \left(\chi_1^2 - \chi_2^2\right) R - \frac 12 \left(\chi_1\nabla^2\chi_1 - \chi_2\nabla^2\chi_2\right) + U \left(\chi_1, \chi_2 \right) ~.
\end{equation}
Note that one of the scalar fields, $\chi_2$, in this Lagrangian has the correct sign in front of its kinetic part. Similar model without the potential $U$ was considered in \cite{McFadden:2004se}.


\subsection{Models with Non-Minimal Couplings}

We can choose a potential $U$ that preserves the conformal symmetry and which leads to
\begin{equation}
L_G = \xi \left(\chi_1^2 - \chi_2^2\right) R - \frac 12 \left(\chi_1\nabla^2\chi_1 - \chi_2\nabla^2\chi_2\right) + u_1\chi_2^{10/3} + u_2\chi_2^{-2/3}\chi_1^{4}~,
\end{equation}
where we can fix $\chi_2 = {\rm constant} = c$. If we identify
\begin{equation}
\xi c^2 = \frac 12~, \qquad u_1c^{10/3} = \Lambda~, \qquad u_2c^{-2/3} = \lambda~,
\end{equation}
we finally get
\begin{equation}\label{Nonminimal}
L_G =\frac{1}{2}R -\xi\chi_1^2 R - \frac 12 \chi_1\nabla^2\chi_1 + \Lambda+\lambda\chi_1^4~.
\end{equation}
This is the Lagrangian used in \cite{Farakos:2005hz, Farakos:2006tt}, where a smooth numerical solution was found. It was demonstrated that the model is stable and can be used as the background for the construction of a realistic brane-world scenario.


\subsection{Models With Hyperbolic Potentials}

Beyond the conformal symmetry, the Lagrangian (\ref{SO(1,1)}) has a $SO(1,1)$ internal symmetry if $U = 0$. The non-zero potential that preserves these symmetries is given by
\begin{equation}
U = u \left(\chi_1^2 - \chi_2^2 \right)^{5/3}~.
\end{equation}
As above, we can use the conformal symmetry to fix the gauge
\begin{equation}\label{gauge}
\xi \left(\chi_1^2 - \chi_2^2 \right) = 2M^3
\end{equation}
and the gravitational part of (\ref{SO(1,1)}) reduces to the standard Einstein-Hilbert Lagrangian. We also get that the potential reduces to a cosmological constant $U = \Lambda$. This constraint also provides that the model has the solutions:
\begin{equation}
\begin{split}
\chi_1 &= \sqrt{\frac {2M^3}{\xi}} ~\cosh \left(\sqrt{\frac{\xi}{2M^3}}\phi \right)~, \\
\chi_2 &= \sqrt{\frac {2M^3}{\xi}} ~\sinh \left(\sqrt{\frac{\xi}{2M^3}}\phi \right) ~.
\end{split}
\end{equation}
With this, the Lagrangian (\ref{SO(1,1)}) becomes
\begin{equation}
L_G = M^3 R + \frac 12 \phi\nabla^2\phi + \Lambda ~.
\end{equation}
An important point about this Lagrangian is that we get a correct sign for the kinetic part of the scalar field and also that we have a cosmological constant. However, this do not provide a kink-like solution for a braneworld. For this we must break the $SO(1,1)$ symmetry. Preserving the conformal symmetry, this can be done if we change our potential $U$ by a dimensionless function $V \left(\chi_1,\chi_2 \right)$, namely
\begin{equation}\label{break}
\tilde{U} = u \left(\chi_1^2 - \chi_2^2\right)^{5/3} V \left(\chi_1,\chi_2 \right)~.
\end{equation}
If $V = 1$ we recover our symmetry. Using the above solution, we finally get for our effective potential
\begin{equation}
\tilde{U} = u \left( \chi_1^2 - \chi_2^2 \right)^{5/3} F\left(\sinh \sqrt{\frac{\xi}{2M^3}}\phi, \cosh \sqrt{\frac{\xi}{2M^3}}\phi\right) ~.
\end{equation}
Note that this kind of potential can be naturally generated by fixing the energy scale of gravity in 5D.

Since the function $V \left(\chi_1,\chi_2 \right)$ is arbitrary, we can generate from it many hyperbolic potentials commonly used in the literature. The equations of motion are given by
\begin{equation}
\begin{split}\label{coupledeqSO(1,1)}
\frac 12 \phi'^2 - V(\phi) - \Lambda &= 24M^3 A'^2 ~, \\
\frac 12 \phi'^2 + V(\phi) + \Lambda &= - 12M^3A'' - 24M^3 A'^2 ~.
\end{split}
\end{equation}
A solution to this system can again be found by introducing a superpotential $W$ in the form:
\begin{equation}
U(\phi) = \frac 12 \left(\frac{\partial W}{\partial\phi}\right)^2 - \frac{1}{6M^3}W^2 ~.
\end{equation}
It is easy to check that
\begin{equation}
\phi' = \frac {\partial W}{\partial\phi}~, \qquad A' = -\frac {W}{12M^3}
\end{equation}
are solutions to the system (\ref{coupledeqSO(1,1)}). We should note the difference of signs with the case of the phantom scalar field in the last section. However, we have not been able to find explicit solutions for this case.


\section{Conclusions}

In this manuscript we are suggesting to use the hidden conformal symmetry in order to generate smooth Randall-Sundrum type braneworlds by introducing of two scalar fields non-minimally coupled to gravity. Two possibilities were considered.
\begin{itemize}
\item{For the case where both scalar fields are phantom-like, the model has an additional $SO(2)$ symmetry. Then we can obtain two models: a) the Standing Wave Braneworld \cite{Gogberashvili:2009yp, Gogberashvili:2010yf, Gogberashvili:2016vwt} and b) the Sine-Gordon model with negative tension brane \cite{Koley:2004au}.}

\item{The model with one phantom and one normal scalar fields has the additional $SO(1,1)$ symmetry. In this case we get the Randall-Sundrum model with non-minimal coupling \cite{Farakos:2005hz, Farakos:2006tt}. By breaking $SO(1,1)$ we can also obtain a model with effective arbitrary hyperbolic potential.}
\end{itemize}
In general, a symmetry restricts the allowed interactions of the model and can be used as the guide to find the form of a potential that can generate new solutions. For instance, the conformal symmetry provides models with dimensionless parameters, which are important from the quantum viewpoint. Note that the hidden conformal symmetry in models of chaotic inflation is at the center of a prolific production in this area, e.g. the models with complex scalar fields, or generalizations to a K\"{a}ller manifold (see \cite{Kallosh:2013wya, Kallosh:2013xya} and references therein). Despite to the fact that this symmetry has a null Noether current\cite{Jackiw:2014koa,Quiros:2014wda}, it can be used as a solution-generating technique.  A possible generalization is to consider the Weyl-integrable geometry (WING) studied in \cite{Quiros:2014wda}. Therefore, the method introduced in this paper is the starting point for the generalizations and constructions of smooth brane models.


\section*{Acknowledgments}

The authors would like to thank Alexandra Elbakyan for removing all barriers in the way of science.

G. Alencar acknowledges the support of Funda\c{c}\~ao Cearense de Apoio ao Desenvolvimento Cient\'ifico e Tecnol\'ogico (FUNCAP) through PRONEM PNE-0112-00085.01.00/16 and the Conselho Nacional de Desenvolvimento Cient\'ifico e Tecnol\'ogico (CNPq).

M. Gogberashvili acknowledges the support of Shota Rustaveli National Science Foundation of Georgia (SRNSFG) [DI-18-335/New Theoretical Models for Dark Matter Exploration].


\end{document}